\documentclass[%
reprint,
amsmath,amssymb,
aps,
]{revtex4-1}

\usepackage{graphicx}
\usepackage{dcolumn}
\usepackage{bm}
\usepackage[colorlinks,citecolor=blue,urlcolor=blue,linkcolor=blue]{hyperref}

\usepackage[normalem]{ulem}
\usepackage{gensymb}
\usepackage{xcolor}

\begin{document}
	
	\preprint{APS/123-QED}
	
	\title{Pseudospin-orbit coupling and non-Hermitian effects in the Quantum Geometric Tensor of a plasmonic lattice}
	
	\author{Javier Cuerda}
	\email{javier.cuerda@aalto.fi}
	\author{Jani M. Taskinen}
	\author{Nicki K\"allman}
	\author{Leo Grabitz}
	\author{P\"aivi T\"orm\"a}%
	\email{paivi.torma@aalto.fi}
	\affiliation{%
		Department of Applied Physics, Aalto University School of Science, Aalto FI-00076, Finland
	}%

	\date{\today}
	
	\setlength{\parskip}{0pt} 
	
	\begin{abstract}
		We theoretically predict the full quantum geometric tensor, comprising the quantum metric and the Berry curvature, for a square lattice of plasmonic nanoparticles. The gold nanoparticles act as dipole or multipole antenna radiatively coupled over long distances. The photonic-plasmonic eigenfunctions and energies of the system depend on momentum and polarization (pseudospin), and their topological properties are encoded in the quantum geometric tensor. By T-matrix numerical simulations, we identify a TE-TM band splitting at the diagonals of the first Brillouin zone, that is not predicted by the empty lattice band structure nor by the highly symmetric nature of the system. Further, we find quantum metric around these regions of the reciprocal space, and even a non-zero Berry curvature despite the trivial lattice geometry and absence of magnetic field. We show that this non-zero Berry curvature arises exclusively from non-Hermitian effects which break the time-reversal symmetry. The quantum metric, in contrast, originates from a pseudospin-orbit coupling given by the polarization and directional dependence of the radiation.

	\end{abstract}
	
	\maketitle
	
	
	\section{\label{sec:Introduction}Introduction}
	
	The search of novel states of matter demands new tools for classifying the underlying structure of eigenstates describing the system in the form of a non-trivial geometry, beyond the traditional classification based on the energy band structure.~Concepts such as the Berry curvature or the Chern number have contributed to this, revealing that the key properties of the system may be protected in the form of topological invariants that take discrete values. The discovery of topological insulators and superconductors \cite{KlitzingPRL1980,ThoulessPRL1982,BernevigSuperconductors} has recently been extended to bosonic (photonic) systems \cite{PriceJPhysPhot2022,OtaNanophotonics2020,OzawaRevModPhys2019,KhanikaevNatPhot2017,SunProgQuantElec2017,LuNatPhot2014,SmirnovaAPR2020,WangNature2009,TangLPR2022,RechtsmanNature2013,KhanikaevNatMat2013,SolnyshkovOptMatExp2021} that may break the time-reversal symmetry~\cite{NagaosaRevModPhys2010} and feature, as the former, chiral edge states and protection against imperfections~\cite{HatsugaiPRL1993}.  
	
	The quantum geometric tensor (QGT)~\cite{Provost1980} has become an important concept as it provides the structure of eigenfunctions (Bloch states) of a Hamiltonian. Its real part is the quantum metric~\cite{PeottaNatComm2015,LiangPRB2017,HuhtinenPRB2022,TormaNatRevPhys2022,PiechonPRB2016,BraunRevModPhys2018}, related to the distance between eigenstates, and the imaginary part gives the Berry curvature~\cite{Berry1984,NagaosaRevModPhys2010} that characterizes their phase. Recently, the full QGT has been measured in several systems~\cite{ZhengChinPhysLett2022,YuNatSciRev2019,YiArxiv2023,GianfrateNature2020,RenNatComm2021}. Further, non-Hermitian effects \cite{ElGanainyNatPhys2018,ZhaoScience2019,Moiseyev2011,BergholtzRevModPhys2021,DingNatRevPhys2022,OkumaAnnuRevCondMat2023,NasariOptMatExp2023,LiuNanophotonics2023,WangJOSAB2023,GongPRX2018,KawabataPRX2019,YaoPRL2018,DopplerNature2016,ChenNature2017,WiersigPRL2014,MiriScience2019,OzdemirNatMat2019,DennerNatComm2021,ShenPRL2018,LeykamPRL2017,SuSciAdv2021} have been shown to manifest as non-trivial quantum geometric phenomena~\cite{SolnyshkovPRB2021}, and very recently non-Hermitian quantum metric has been reported~\cite{LiaoPRL2021}. In Ref.~\cite{CuerdaPRL2023}, we experimentally show the existence of a non-zero quantum metric, and the pioneering observation of non-Hermitian Berry curvature in a plasmonic lattice. 
	
	Both in electronic and photonic systems, Berry curvature can be created by breaking the time reversal or some other symmetry~\cite{Vanderbilt2018}. Moreover, even in time-reversal invariant systems, the spin-orbit coupling can generate an effective magnetic field that produces non-trivial quantum geometric or even topological phenomena~\cite{Kane1PRL2005,Kane2PRL2005}. The spin-orbit coupling can be induced through several mechanisms: for example, it can come from a genuine magnetic field interacting with spin up and spin down states \cite{BernevigSuperconductors}, or from fluxes introduced by complex hoppings in a two-site unit cell lattice~\cite{HaldanePRL1988}, or from a specific light polarization (pseudospin) dependence of the band structure of a photonic system~\cite{KavokinPRL2005}. Here we unveil, with detailed numerical simulations, the connection of such phenomena with the observations in Ref.~\cite{CuerdaPRL2023}. In particular, we find that square plasmonic lattices have a pseudospin-orbit coupling originating from their specific radiation properties, and that time-reversal-symmetry breaking by losses plays a crucial role in the quantum geometric properties of the system.
	
	The structure of this article is the following. 
	We review the definition of the QGT in Section \ref{sec:qgtdefs}. The quantum geometric tensor describes properties of the single particle wavefunction, it does not involve multiparticle and interactions effects which could lead to entanglement or quantum statistics effects. Thus, our results apply for both quantum and classical regimes. The experiments in Ref.~\cite{CuerdaPRL2023} were performed in the many-photon (classical) setting. 
	In Section \ref{sec:tmatrix_eigenf} we implement a transition (T-)matrix approach that accounts for both the long-range radiative interactions within a plasmonic lattice and the optical response of the individual metallic nanoparticles. By means of infinite summation techniques across the whole lattice, the collective modal dispersion relations $\omega'(\mathbf{k})$ and the losses of each mode $\omega''(\mathbf{k})$ are numerically obtained in the reciprocal space. With this method, we investigate the possible removal of degeneracies and gap openings that typically lead to non-trivial topological phenomena \cite{HaldanePRL2008,RaghuPRA2008}. Here we reveal a TE-TM band splitting along the diagonals of the first Brillouin zone, for both the real (band energy) and the imaginary part (modal losses) of the eigenfrequency. 
	In Section \ref{sec:qgt_tmatrix}, by systematically tracking the eigenfunctions obtained with the T-matrix approach across a grid in $k-$space, we explicitly calculate the quantum metric and Berry curvature. Our results show a clear non-zero contribution of all the components of the QGT along the diagonals of the Brillouin zone, intimately related to band degeneracy removal and dissipation losses. Finally, in Section~\ref{sec:lossdepQGT}, we discuss on the nature of time-reversal symmetry breaking by losses (or gain) and its relation with the Berry curvature.

	\section{\label{sec:qgtdefs}Quantum geometric tensor: definitions}
	Here we review basic concepts of band geometry and topology that are utilized throughout this paper. We consider systems that are periodic in real space, thus the eigenmodes of the system $|u_{n,\mathbf{k}}\rangle$ correspond to Bloch bands with energies $E_{n,\mathbf{k}}$, where $n$ is the band index and $\mathbf{k}$ is a vector in the reciprocal space. We will consider lattice modes that are confined in the plane of the lattice, hence we may restrict our treatment to the two-dimensional $k-$space: $\mathbf{k}=(k_{x},k_{y})$. All the definitions used, as well as our theoretical results, apply for quantum and classical wave functions equally. We use the terminology ``quantum geometric tensor'' and ``quantum metric'' as a convention, even if they can be used to characterize classical systems/parameter regimes too.
	
	We investigate the underlying quantum geometry that describes small changes in the parameters $\mathbf{k}\rightarrow\mathbf{k}'=\mathbf{k}+d\mathbf{k}$ that affect $|u_{n,\mathbf{k}}\rangle$. To this purpose, we first introduce the quantum metric, also called Fubini-Study metric.  We start from the definition of the infinitesimal distance between the states $|u_{n,\mathbf{k}}\rangle$ and $|u_{n,\mathbf{k}'}\rangle$:
	\begin{equation}\label{fubinistudy}
		ds^{2}=1-|\langle u_{n,\mathbf{k}}|u_{n,\mathbf{k}+d\mathbf{k}}\rangle|^{2}.
	\end{equation}
	By means of Taylor expansion to second order in Eq.~\eqref{fubinistudy}, combined with the normalization $\langle u_{n,\mathbf{k}}|u_{n,\mathbf{k}}\rangle=1$, the metric that defines such a distance is characterized by a real, second-rank symmetric tensor $ds^{2}=g_{ij}^{n}dk_{i}dk_{j}$, defined as follows:
	\begin{align}\label{def:quantummetric}
		g_{ij}^{n}&=\Re\bigg\{\bigg\langle\dfrac{\partial u_{n,\mathbf{k}}}{\partial k_{i}}\bigg|\dfrac{\partial u_{n,\mathbf{k}}}{\partial k_{j}}\bigg\rangle\nonumber\\
		&\qquad\qquad-\bigg\langle\dfrac{\partial u_{n,\mathbf{k}}}{\partial k_{i}}\bigg|u_{n,\mathbf{k}}\bigg\rangle\bigg\langle u_{n,\mathbf{k}}\bigg|\dfrac{\partial u_{n,\mathbf{k}}}{\partial k_{j}}\bigg\rangle\bigg\}.
	\end{align}
	The quantum metric \eqref{def:quantummetric} is gauge-invariant, hence a measurable property of the system.
	
	Another important gauge-invariant quantity is the Berry curvature. While the quantum metric is derived from the modulus of two quantum states via the distance~\eqref{fubinistudy}, the Berry curvature can be obtained by using the phase from the complex inner product of states $\langle u_{n,\mathbf{k}}|u_{n,\mathbf{k}' }\rangle=re^{i\Delta\varphi_{\mathbf{k},\mathbf{k}'}}$. By iteratively applying this operation along a closed loop in $k-$space, it is found that the accumulated phase only depends on the followed geometrical path $\gamma$ \cite{Vanderbilt2018}. This is the celebrated Berry phase, and reads as follows:
	\begin{equation}\label{def:berryphase}
		\varphi_{B}=i\oint_{\gamma}\langle u_{n,\mathbf{k}}|\nabla_{\mathbf{k}}u_{n,\mathbf{k}}\rangle\cdot d\mathbf{k}.
	\end{equation}
	Using the Stokes theorem, we may rewrite Eq.~\eqref{def:berryphase} as a surface integral, obtaining:
	\begin{equation}\label{berryphase_surface}
		\varphi_{B}=\int_{S}\sum_{i<j}\mathfrak{B}_{ij}^{n} dk_{i}\wedge dk_{j},
	\end{equation}
	where $S$ is the oriented surface enclosed by the contour $\gamma$. The integrand of Eq.~\eqref{berryphase_surface} is the Berry curvature:
	\begin{equation}\label{def:berrycurvature}
		\mathfrak{B}_{ij}^{n}=-2\Im\bigg\langle\dfrac{\partial u_{n,\mathbf{k}}}{\partial k_{i}}\bigg|\dfrac{\partial u_{n,\mathbf{k}}}{\partial k_{j}}\bigg\rangle,
	\end{equation}
	which is an antisymmetric second-rank tensor independent of the choice of $\gamma$ and $S$. The gauge-invariant quantum metric~\eqref{def:quantummetric} and Berry curvature~\eqref{def:berrycurvature} tensors characterize the geometry of the quantum state manifold, and may be combined into the quantum geometric tensor~\cite{Provost1980}:
	\begin{align}\label{def:qgt}
		T_{ij}^{n}&=\Big\langle\dfrac{\partial u_{n,\mathbf{k}}}{\partial k_{i}}\Big|\dfrac{\partial u_{n,\mathbf{k}}}{\partial k_{j}}\Big\rangle\nonumber\\
		&\qquad\qquad-\Big\langle\dfrac{\partial u_{n,\mathbf{k}}}{\partial k_{i}}\Big| u_{n,\mathbf{k}}\Big\rangle\Big\langle u_{n,\mathbf{k}}\Big|\dfrac{\partial u_{n,\mathbf{k}}}{\partial k_{j}}\Big\rangle.
	\end{align}
	The real and imaginary parts of the QGT in Eq.~\eqref{def:qgt} relate directly to the quantum metric and Berry curvature: $g_{ij}^{n}=\Re T_{ij}^{n}$, and $\mathfrak{B}_{ij}^{n}=-2\Im T_{ij}^{n}$.
	
	Although expressions~\eqref{def:quantummetric} and~\eqref{def:qgt} have been applied previously for studies of non-Hermitian systems \cite{SolnyshkovPRB2021}, it should be noted that they are based on the hermiticity of the norm: $\langle u_{n,\mathbf{k}}|u_{n,\mathbf{k}}\rangle=1$. It has been argued that the non-Hermitian version of the quantum metric is more general than Eq.~\eqref{def:quantummetric} and can be defined via the left and right eigenstates \cite{HePRA2021}. Moreover, for strong non-hermiticity, bands may become degenerate and coalesce into an exceptional point \cite{SolnyshkovPRB2021}. In such a case, the winding number of a complex effective field replaces the Chern number as the meaningful invariant~\cite{,LeykamPRL2017}. In this work, we study non-degenerate bands (in fact, degeneracy is lifted due to losses, see Section \ref{sec:tmatrix_eigenf}), suggesting that the non-hermiticity is weak \cite{ShenPRL2018}. Further, we show in Section \ref{sec:qgt_tmatrix} that the Berry curvature of non-Hermitian origin is several orders of magnitude smaller than the quantum metric. Since the QGT is a positive semi-definite tensor, the quantum metric is bounded from below by the Berry curvature, $\sqrt{\text{det}(g)}\geq|\mathfrak{B}_{xy}|/2$ \cite{OzawaPRB2021}, and our results are consistent with this. All this suggests that the definitions \eqref{def:quantummetric}, \eqref{def:berrycurvature} and \eqref{def:qgt} are a good approximation for our study due to the weakness of the non-Hermitian effects. The consequences of non-hermiticity are thoroughly examined in Ref.~\cite{CuerdaPRL2023},  by explicitly concerning the left and right eigenstates of the system. 
	
	\section{\label{sec:tmatrix_eigenf}T-matrix simulations and band tracking}
	\begin{figure*}
		\includegraphics[width=0.99\textwidth]{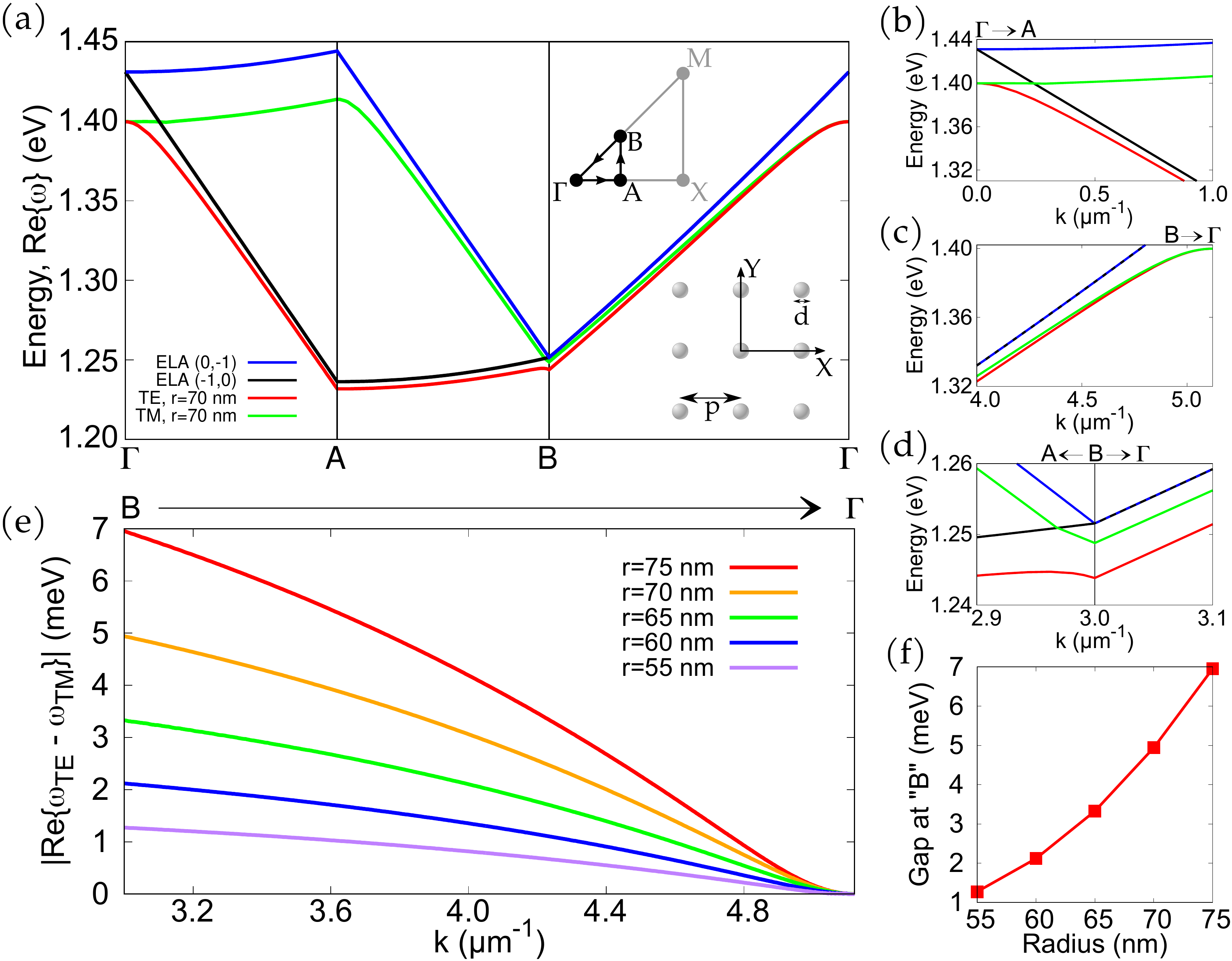}
		\caption{\label{fig:tmatrix_bands}Degeneracy removal of dispersion bands along the diagonal of the first Brillouin zone. (a) Band structure of SLRs that emerge from the empty lattice bands $(-1,0)$ and $(0,-1)$, here called TE and TM bands, respectively. Band energies are given by the real part of the complex eigenfrequency calculated from T-matrix simulations along the path $\Gamma-A-B-\Gamma$ in $k-$space (see inset), where $A=(1.5,0)$ µm$^{-1}$ and $B=(1.5,1.5)$  µm$^{-1}$. A schematic representation of the square lattice is included (only few unit cells are shown) with the typical measures $p=570$ nm and $d=2r$. The radius of the nanoparticles is $r=70$ nm in panels (a)-(d). (b)-(d) Zoomed-in selected areas from panel (a) where bands are degenerate in the empty lattice approximation. While the SLR bands remain degenerate at the $\Gamma$-point, a splitting can be observed along the $B-\Gamma$ trajectory with a gap opening at the point $B$. (e) Energy difference between the TE and TM bands along the trajectory $B-\Gamma$ for various nanoparticle sizes. In (b)-(e), the $k-$axis values denote distance traveled along the $\Gamma-A-B-\Gamma$ path. (f) Size of the energy difference (gap) at $B$ as a function of the nanoparticle radius.
		}
	\end{figure*}
	We calculate the collective response of an infinite square plasmonic lattice, see Fig.~\ref{fig:tmatrix_bands}(a), including the full optical description of the individual metallic nanoparticles as well as the radiative interactions explained in Ref.~\cite{CuerdaPRL2023}. Plasmonic lattices as considered here have eigenmodes that are hybrids of the localized plasmonic resonances and the diffraction orders of the lattice~\cite{ZhouNatNano2013,SchokkerOptica2016,DeGiorgiACSPhot2018,WangMatToday2018,WuNanoLett2020,RamezaniOptica2017,HakalaNatPhys2018,VakevainenNatComm2020,KoshelevACSPhot2021,GuanChemRev2022,CastellanosAdvOptMat2023,RiderACSPhot2022,TaskinenNanoLett2021,HaNatNano2018,MuraiACSPhot2020,HeilmannACSPhot2022,LiuNanophotonics2021,SalernoPRL2022,KravetsChemRev2018,AuguiePRL2008,RodriguezPRX2011,GuoPRB2017,KnudsonACSNano2019,WeickPRL2013,KuznetsovScience2016,KrasnokIEEE2020}. These modes are called surface lattices resonances (SLRs); for their description within the empty lattice approximation, and further information and references, see Ref.~\cite{CuerdaPRL2023}.
	
	Most of the previous theoretical studies describing the effect of long-range interactions and retardation treated the metallic nanoparticles as point dipoles \cite{DowningPRL2019,MartikainenPTRSA2017,PocockACSPhot2018}. We utilize an open source implementation \cite{Necada2021,QPMScode} of the more general transition matrix (T-matrix) method \cite{Mishchenko2000}, that accounts for higher multipolar orders that may become relevant for large particle sizes beyond the dipole approximation. We numerically extract the eigenfunctions corresponding to the modes of the system, which is crucial for the QGT calculation in Eq.~\eqref{def:qgt}. Within this approach, the lattice modes are governed by the following expression:
	\begin{equation}\label{tmatrixeq}
		(I-TW)\mathbf{f}(\mathbf{k})=0.
	\end{equation}
	Here, $T$ is the transition matrix that describes the individual nanoparticles modelled as metallic nanospheres, $W$ carries out an infinite Ewald summation across every lattice site, $I$ is the identity matrix, and $\mathbf{f}(\mathbf{k})$ is a column vector that contains complex coefficients. Detailed derivation and definitions are found in Appendix \ref{appendix:tmatrix} and Ref.~\cite{Necada2021}. In the T-matrix method, the electric field is expanded in a basis of vector spherical wavefunctions (VSWFs), labeled with subindices $\{\tau,l,m\}$ that correspond to electric ($\tau=2$) and magnetic ($\tau=1$) type of solutions to the Maxwell equations, as well as the multipole degree $l=1,2,\ldots$ corresponding to dipolar, quadrupolar, and higher multipole orders. For instance, the electric field from a scatterer in one unit cell is given by the summation:
	\begin{equation}\label{scatfieldtmatrix}
		\mathbf{E}_{sc}(\omega,\mathbf{r})=\sum_{\tau=1,2}\sum_{l=1}^{\infty}\sum_{m=-l}^{l}f_{\tau lm}\mathbf{u}_{\tau lm}(\kappa(\mathbf{r})),
	\end{equation}
	where $\mathbf{u}_{\tau lm}(\kappa(\mathbf{r}))$ are called \emph{outgoing} VSWFs \cite{Kristensson2016}. The components of the coefficient vector $\mathbf{f}$ in Eq.~\eqref{tmatrixeq} are the coefficients $f_{\tau lm}$ in Eq.~\eqref{scatfieldtmatrix} that determine the eigenmodes of the system. We thus define our quantum state $|u_{\mathbf{k}}\rangle\equiv \mathbf{f}(\mathbf{k})$ as the column vector with coefficients that follow: 
	\begin{align}\label{fcoefstaulm}
		f_{\tau l m}\in\mathbb{C}\ |\quad &\tau=1,2;\ l=1,2,\ldots,l_{max};\\
		&m=-l,-l+1,\ldots,+l\nonumber.
	\end{align}
	The length of the eigenvector $|u_{\mathbf{k}}\rangle$ in the space of VSWFs is defined by the sum in Eq.~\eqref{scatfieldtmatrix}, and ideally ranges up to $l_{max}=\infty$ in (\ref{fcoefstaulm}). In our implementation, we set a cutoff to $l_{max}$ and truncate the sum of Eq.~\eqref{scatfieldtmatrix}, such that elements $T_{\tau lm}^{\tau' l'm'}$ of the $T$ matrix are considered negligible for $l,l'\geq l_{max}$ \cite{Necada2021}. The length of the eigenvector is given by $N_{m}\equiv2l_{max}(2l_{max}+1)$ and accordingly the dimension of the square matrix $(I-TW)$ is $N_{m}^{2}$. In this work, $l_{max}=2$ was fixed, and convergence of the results was tested with higher values.
	
	Solving the problem of Eq.~\eqref{tmatrixeq} involves calculating both eigenvalues and eigenvectors of the matrix $M(\omega)\equiv I-TW$. Even at the individual particle level, the T-matrix is non-Hermitian for lossy scatterers, and in general the eigenvalues of the full problem matrix $M(\omega)$ are complex eigenfrequencies of the form $\omega=\omega'+i\omega''$, where $\omega',\omega''\in\mathbb{R}$. The real part $\omega'$ of the obtained eigenfrequency $\omega$ gives the energy of the collective lattice mode and the imaginary part $\omega''$ accounts for both the radiative and dissipative losses of the mode. Our numerical procedure calculates the complex eigenvalues $\omega(\mathbf{k})$ for each value of the momentum parallel to the lattice plane $\mathbf{k}=(k_{x},k_{y})$. To that purpose, we implement Beyn's integral method that numerically locates all the complex eigenvalues of the matrix problem (\ref{tmatrixeq}) within a defined contour in the complex frequency plane. Eigenvectors corresponding to each eigenvalue are then obtained, see Refs.~\cite{Necada2021} and \cite{Beyn2012} for details.
	
	We first study the band dispersion energies of the SLR modes, see Fig.~\ref{fig:tmatrix_bands}(a).~Given a linear path in the two-dimensional $k-$space between generic points $\mathbf{k}_{0}$ and $\mathbf{k}_{f}$, discretized by $N_{k}$ points such that $\mathbf{k}_{j}=\mathbf{k}_{0}+j(\mathbf{k}_{f}-\mathbf{k}_{0})/N_{k}$ with $j=0,1,\ldots,N_{k}$, we consecutively determine the eigenvalues $\omega_{i}(\mathbf{k}_{j})$, $i=1,\ldots,N \leq N_{m}$ inside an elliptical contour in the complex frequency plane, for each $\mathbf{k}_{j}$. We focus on a range of energies around the first $\Gamma$-point by tracking the path $\Gamma-A-B-\Gamma$ within the irreducible Brillouin zone (see the inset of Fig.~\ref{fig:tmatrix_bands}(a)). Studies of other high-symmetry points is an interesting direction for future work \cite{GuoPRL2019,JuarezACSPHot2022}.
	
	Since many physical solutions result from the calculation, some prior knowledge of SLRs in a square lattice is instrumental for identifying the nature of each eigenmode. A good guide to the dispersions of SLRs is provided by the empty lattice approximation (see Ref.~\cite{CuerdaPRL2023}) where four diffraction orders cross at the $\Gamma-$point. Along the high symmetry line $\Gamma-X$ in Fig.~\ref{fig:tmatrix_bands}(a), modes can be classified according to their polarization properties: assuming that the nanoparticles behave dominantly as electric dipoles polarized in the same direction as the incoming light, the modes $(\pm1,0)$ can only be excited with transverse electric (TE, $y$-polarized) radiation, likewise, $(0,\pm1)$ are transverse magnetic (TM) polarized modes \cite{GuoPRB2017}. We keep this naming for SLR bands in Fig.~\ref{fig:tmatrix_bands}, and even along other trajectories where modes are not purely TE or TM polarized.
	
	Scattering by the nanoparticles couples the diffraction orders at the $\Gamma-$point and forms two TE SLR modes separated in energy by a small bandgap \cite{RodriguezPRX2011,SchokkerPRB2014,SchokkerPRB2017}. As a consequence, one of the TE modes is shifted to lower energies than the empty lattice approximation (red line in Fig.~\ref{fig:tmatrix_bands}(a)), while the other is shifted to higher energies (not shown in Fig.~\ref{fig:tmatrix_bands}(a)). The TM modes (green line in Fig.~\ref{fig:tmatrix_bands}(a)) remain degenerate with both the TE mode at the $\Gamma-$point, and with each other along the high-symmetry line $\Gamma-A$. In what follows, we focus on the TE SLR that forms from the $(-1,0)$ empty lattice mode, and on the TM mode coming from $(0,-1)$.
	
	The bandgap opening at the $\Gamma-$point and other high symmetry points occurs because several diffraction orders cross in those points. Away from the high symmetry points the dispersions are expected to resemble those of the empty lattice approximation \cite{GuoPRB2017}. 
	Indeed this is the case around the point $A$ in Fig.~\ref{fig:tmatrix_bands}(a). However, a different behavior is found around the $B-$point of Fig.~\ref{fig:tmatrix_bands}(a), placed on the diagonal of the first Brillouin zone. Tracking the bands in $k-$space along the diagonal $B-\Gamma$ reveals that the TE and TM dispersion energies do not remain degenerate as in the empty lattice approximation. Instead, the degeneracy is lifted increasingly while moving away from the $\Gamma-$point along the diagonal. To further examine this behavior, we provide in Figs.~\ref{fig:tmatrix_bands}(b)-(d) zoomed-in areas of the SLR bands studied in Fig.~\ref{fig:tmatrix_bands}(a).
	
	Fig.~\ref{fig:tmatrix_bands}(b) shows that the degeneracy of the TE (red line) and TM bands (green line) holds at the $\Gamma$-point for both the empty lattice approximation and realistic simulations with a finite particle radius of $r=70$ nm. TE and TM degeneracy at the $\Gamma-$point is understood from the $x-y$ symmetry of the square lattice, as the system is invariant under $\pi/2$ rotations around the out-of-plane $z$-axis, regardless of the particle size. Away from the $\Gamma$-point along the trajectory $\Gamma-A$ of Fig.~\ref{fig:tmatrix_bands}(b), the $x$-$y$ symmetry is broken with $k_{x}\neq0$, thus degeneracy is lifted and the TE and TM modes emerge. Energy loss of photons by scattering with the metallic nanoparticles explains the TE and TM energy bands departing further from the empty lattice approximation in Fig.~\ref{fig:tmatrix_bands}(b). However, degeneracy of these bands at the $\Gamma$-point is not lifted by the presence of the nanoparticles.
	
	A different scenario is presented in Figs.~\ref{fig:tmatrix_bands}(c),(d), as the presence of nanoparticles clearly lifts the empty lattice degeneracy along the trajectory $B-\Gamma$. Fig.~\ref{fig:tmatrix_bands}(c) shows that, while the degeneracy exactly at the $\Gamma-$point is respected consistently with Fig.~\ref{fig:tmatrix_bands}(b), finite sized particles remove the degeneracy of TE and TM bands along the diagonal of the first Brillouin zone. As a consequence, a gap is created between these bands at the point $B$ of the chosen path (see Fig.~\ref{fig:tmatrix_bands}(d)). We achieve insight into the degeneracy removal at the diagonal $B-\Gamma$ by tracking the energy difference between the TE and TM dispersion bands. Fig.~\ref{fig:tmatrix_bands}(e) shows that the band splitting increases with the particle size and is ever growing along this trajectory when moving away from the $\Gamma-$point. Interestingly, a superlinear dependence of the gap at the point $B$ is found as a function of the particle radius, see Fig.~\ref{fig:tmatrix_bands}(f). This implies that the splitting can be experimentally detected with a moderate particle size (e.g., we find an energy difference of 7~meV with a particle radius of 75~nm). We note that $B$ is not a high-symmetry point of the square lattice, and is introduced only to illustrate the degeneracy removal along the diagonal of the first Brillouin zone.
	
\begin{figure*}
		\includegraphics[width=0.99\textwidth]{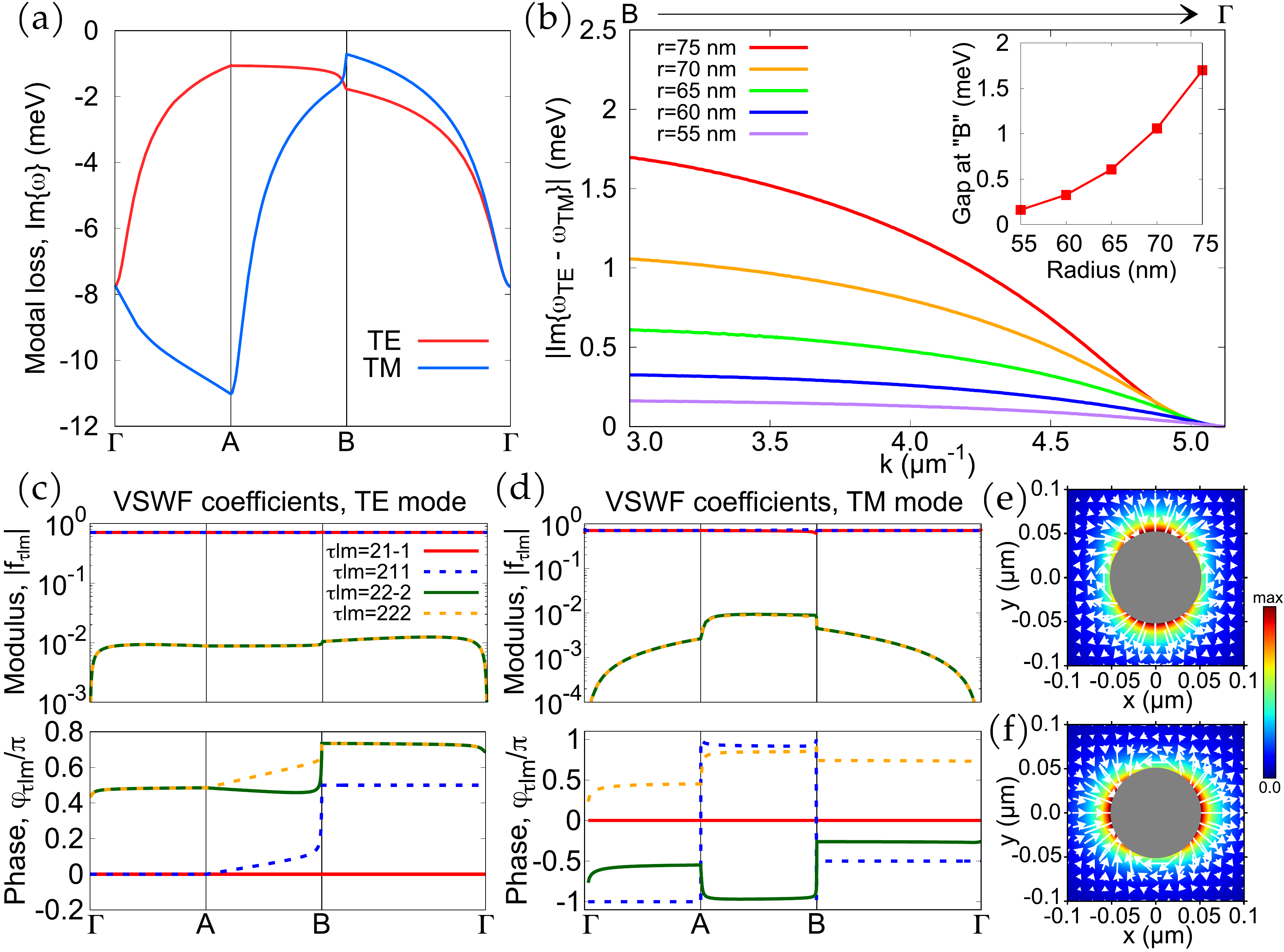}
		\caption{\label{fig:losses_eigenvecs}Modal losses, eigenvectors, and field profiles of the TE and TM SLR modes shown in Fig.~\ref{fig:tmatrix_bands}. (a) Modal losses along the path $\Gamma-A-B-\Gamma$ within the first Brillouin zone, calculated as the imaginary part of the mode eigenfrequency from T-matrix simulations of a lattice of nanospheres with radius $r=70$ nm. (b) Losses of the TE mode subtracted from those of the TM mode along the trajectory $B-\Gamma$ for various particle sizes. Compared to the dispersion energies in Fig.~\ref{fig:tmatrix_bands}(e), the splitting is smaller yet has a similar trend, with zero loss difference at the $\Gamma$-point, and a gap opening at the point $B$. The inset shows the size of the gap in the imaginary part of the TE and TM eigenfrequencies at the point $B$, as a function of the particle radius. (c) Tracking of the complex coefficients $f_{\tau lm}$ that enter into the eigenvector components of the TE mode: both the modulus (upper panel) and phase (lower panel) of each relevant coefficient are shown. (d) Same as (c), but for the TM mode. Both panels (c) and (d) are calculated for a lattice of metallic nanospheres of radius $r=50$ nm. Eigenvectors obtained for other particle sizes reported in this work are qualitatively similar. (e) Field profile of the TE mode in a unit cell of the lattice (the period is 0.57 µm, but only the neighboring area to the nanoparticles is shown), as calculated at the $\Gamma$-point from the in-phase superposition of VSWFs $\mathbf{u}_{21\pm1}$ according to panel (c).~White arrows indicate both magnitude and direction of the electric field $\mathbf{E}$, and the colorscale map shows its modulus $|\mathbf{E}|$. (f) Field profile of the TM mode at the $\Gamma$-point resulting from the superposition of VSWFs $\mathbf{u}_{21\pm1}$ with a phase difference of $\pi$, as in panel (d).}
	\end{figure*}

	To understand possible non-Hermitian effects, we now focus on the losses inherent to the system.  The losses featured by the TE and TM modes of the plasmonic lattice are given by the imaginary part of the eigenfrequency $\omega''$ obtained from the T-matrix simulations. Fig.~\ref{fig:losses_eigenvecs}(a) shows the evolution of these modal losses for a lattice of particles with radius $r=70$ nm, tracked along the same path in $k-$space as in Fig.~\ref{fig:tmatrix_bands}(a). While the TE and TM dispersion bands in Fig.~\ref{fig:tmatrix_bands}(a) converge to the empty lattice approximation for decreasing particle size, losses in Fig.~\ref{fig:losses_eigenvecs}(a) converge to zero. This is consistent since dissipation within the metal is the main loss mechanism of plasmonic nanoparticles, and the empty lattice approximation deals ideally with lossless photonic modes. We have found that, along the studied path in reciprocal space, the loss-dependence of the TE mode on $\mathbf{k}$ is qualitatively similar regardless of the particle size, and the same applies to the TM mode. However, Fig.~\ref{fig:losses_eigenvecs}(a) shows that the dependence of the TE mode along the path $\Gamma-A-B$ is different from that of the TM mode.
	
	Notably, along the trajectory $B-\Gamma$, the features of the band dispersions in Fig.~\ref{fig:tmatrix_bands}(a)-(d) and modal losses in Fig.~\ref{fig:losses_eigenvecs}(a) are similar: they become equal at the $\Gamma-$point and depart further when moving along the diagonal of the Brillouin zone. In Fig.~\ref{fig:losses_eigenvecs}(b) we show the difference in modal losses between the TE and TM modes, revealing that the trend is analogous to that followed by the band energy difference in Fig.~\ref{fig:tmatrix_bands}(e). Consequently at the point $B$, a gap opening takes place also in the imaginary part of the eigenfrequency. The size of the gap increases with the radius of the nanoparticles (see inset of Fig.~\ref{fig:losses_eigenvecs}(b)), which is consistent with the increased ohmic losses. The splitting of the imaginary eigenfrequency $\omega''$ is smaller than the one found for the real part $\omega'$ at every point on the diagonal $B-\Gamma$, and accordingly the size of the gap in Fig.~\ref{fig:losses_eigenvecs}(b) is smaller than in Fig.~\ref{fig:tmatrix_bands}(f). 
	
	We conclude that both the real and imaginary part of the eigenfrequency exhibit a band splitting along the diagonal of the first Brillouin zone, with a larger effect on the dispersion energies. These findings provide a relevant insight into the quantum geometric and topological properties of the system; in particular, the band energy and modal loss splitting at the diagonal of the Brillouin zone suggests that a complex-valued term $\Omega_{y}(\mathbf{k})=\Omega'_{y}(\mathbf{k})+i\Omega''_{y}(\mathbf{k})$, with $\Omega'_{y},\Omega''_{y}\in\mathbb{R}$ and $\Omega''_{y}\ll\Omega'_{y}$ for every $\mathbf{k}$ must be present in the two band Hamiltonian of Ref.~\cite{CuerdaPRL2023}. In agreement with the results presented here, it is found that the real part $\Omega'_{y}$ (band energy) creates the observed quantum metric, and the imaginary part $\Omega''_{y}$ (modal loss) induces non-zero Berry curvature, see sections~\ref{sec:qgt_tmatrix} and \ref{sec:lossdepQGT}.
	
	We next describe the eigenvectors corresponding to the TE and TM bands above, given by the complex-valued coefficients $f_{\tau lm}(\mathbf{k})$ in the VSWF basis as specified in (\ref{fcoefstaulm}). Analysis of the largest coefficients provides understanding of the relevant multipolar contributions to each mode, therefore we first examine the modulus $|f_{\tau lm}(\mathbf{k})|$ along the path $\Gamma-A-B-\Gamma$. While we have shown that the band dispersions and modal losses are highly dependent on $\mathbf{k}$, only four momentum-dependent coefficients are significant for the considered range of particle sizes and carry the same magnitude along the whole considered path, see Figs.~\ref{fig:losses_eigenvecs}(c),(d). The main contribution to the eigenvector comes from the superposition of two terms of electric dipolar nature, with subindices $\{\tau, l,m\}=\{2,1,1\}$ and $\{2,1,-1\}$ dominating in Fig.~\ref{fig:losses_eigenvecs}(c),(d); the combination of coefficients with $m=\pm 1$ reveals that the modes are polarized in the lattice plane \cite{Necada2021}. A minor contribution comes from the coefficients with subindices $\{\tau, l,m\}=\{2,2,2\}$ and $\{2,2,-2\}$, corresponding to electric quadrupolar components with in-plane polarization.~The typical size of the nanoparticles in our study is relatively large compared to the lattice period~\cite{MartikainenPTRSA2017}. Nevertheless, we have found that the collective behavior of the lattice is governed by modes of electric dipolar nature.
	
	Although the same combinations of $\{\tau, l,m\}$ contribute to the eigenvectors of both the TE and TM modes, the phases $\varphi_{\tau lm}(\mathbf{k})$ of the corresponding coefficients $f_{\tau lm}=|f_{\tau lm}|e^{i\varphi_{\tau lm}}$ differ substantially, see the lower panels of Figs.~\ref{fig:losses_eigenvecs}(c),(d). We choose the overall phase factor of both eigenvectors such that the coefficient $f_{21-1}$ is real along the considered path ($\varphi_{21-1}=0$), and the rest of the relative phases are defined accordingly. Hence the phase $\varphi_{211}$ (blue dashed line in lower panels of Figs.~\ref{fig:losses_eigenvecs}(c),(d)) accounts for the phase difference between the two dominant dipolar VSWFs: $\Delta\varphi\equiv\varphi_{211}-\varphi_{21-1}$. This phase difference is $\Delta\varphi=0$ for the TE mode at the $\Gamma-$point (see Fig.~\ref{fig:losses_eigenvecs}(c)), and then evolves smoothly with $\mathbf{k}$; however, at $B$ (that is, the point placed at the diagonal of the first Brillouin zone) there is an abrupt shift to $\Delta\varphi\approx\pi/2$. Similarly, the phase difference of the TM mode in Fig.~\ref{fig:losses_eigenvecs}(d) is $\Delta\varphi=-\pi$ at the $\Gamma-$point, but experiences a sudden change to $\varphi=-\pi/2$ at the point $B$. Additional features at points $A$ and $B$ in Fig.~\ref{fig:losses_eigenvecs}(d) are attributed to the choice of $(-\pi,\pi]$ as the principal branch of the phase. We further discuss the implications of the behavior of eigenvectors close to the diagonal in Section~\ref{sec:qgt_tmatrix}.
	
	Finally, we use the modulus and phase of the VSWF coefficients that enter into the eigenvector to achieve further insight into the considered modes. 
	Following Eq.~\eqref{scatfieldtmatrix} and disregarding the quadrupolar terms, we may write the field distribution of the modes as the superposition of two outgoing VSWFs:
	\begin{equation}\label{vswf_superposition_dipolar}
		\mathbf{E}=f_{211}\mathbf{u}_{211}+f_{21-1}\mathbf{u}_{21-1}.
	\end{equation}
	At the $\Gamma$-point, we have found that the coefficients $f_{21\pm1}$ have the same modulus with a phase difference. Therefore we may write: $f_{21-1}=\exp(i\Delta\varphi/2)$ and $f_{211}=\exp(-i\Delta\varphi/2)$, where $\Delta\varphi=0$ for the TE mode, and $\Delta\varphi=\pi$ for the TM mode. Figs.~\ref{fig:losses_eigenvecs}(e),(f) show the resulting field profiles in one unit cell of the lattice according to Eq.~\eqref{vswf_superposition_dipolar} that confirm the dipolar character of the analyzed modes. The polarization properties are also revealed: the profile in Fig.~\ref{fig:losses_eigenvecs}(e) can only be created with $y$-polarized light, therefore the mode is TE for propagation in the $x$-direction. By the same reasoning, the mode in Fig.~\ref{fig:losses_eigenvecs}(f) is TM for the same propagation direction. We also note that at the $\Gamma$-point these two modes are doubly degenerate in energy, and hence, due to the symmetry of the square lattice, the field profiles are equivalent to each other under a $\pi/2$ rotation in real space. Further investigation involving the eigenvectors from the T-matrix approach is carried out in Section~\ref{sec:qgt_tmatrix} for a region of $k$-space around the $\Gamma$-point; in particular, we utilize them to calculate the quantum metric and Berry curvature of TE and TM modes discussed in that Section.
	
	\section{\label{sec:qgt_tmatrix}Calculation of the QGT from T-matrix eigenfunctions}
	We have studied the surface lattice resonance modes sustained by a square lattice, and concluded that modes with different polarization properties $-$ TE and TM modes $-$ show intriguing features along the diagonal of the first Brillouin zone, both in the real and imaginary parts of the eigenfrequency from T-matrix simulations. Importantly, the T-matrix method allows fully characterizing the eigenvectors of the corresponding modes in the VSWF basis, and we have found that the components of the eigenvectors are complex-valued due to both radiative and dissipative losses in the system. In this section, we study whether such losses, in combination with the band structure shown above, enable non-Hermitian effects that explain the degeneracy removal along the diagonal of the Brillouin zone. In particular, since losses can break the time-reversal symmetry, they might lead to interesting effects on the quantum geometry and topology of the system. To explore this, we utilize the T-matrix eigenvectors to calculate the full QGT that comprises the quantum metric and the Berry curvature.
	
\begin{figure*}
		\includegraphics[width=0.99\textwidth]{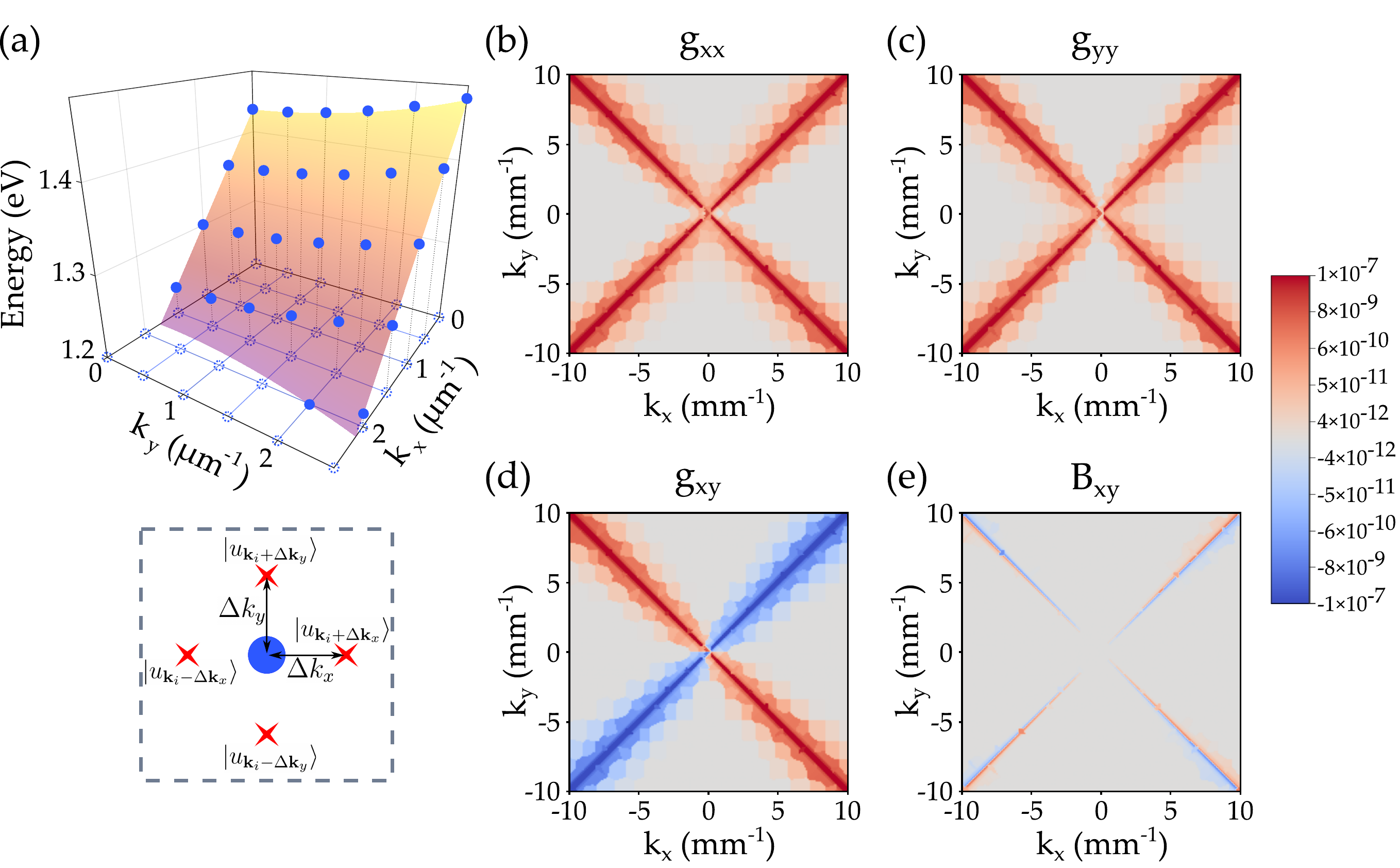}
		\caption{\label{fig:qgt_tmatrix}Quantum geometric tensor (quantum metric and Berry curvature), as obtained from T-matrix simulations of the TE mode in Figs.~\ref{fig:tmatrix_bands} and \ref{fig:losses_eigenvecs}. (a) Schematic representation of the discretization of $k$-space (upper panel with dotted blue circles) and of the numerical procedure that calculates the derivatives of the eigenvectors obtained at each discretization point $\mathbf{k}_{i}$ (lower panel). The calculated eigenmodes (filled blue circles) are shown only for the (-1,0) mode in the first quadrant of the reciprocal space for clarity. Panels (b)-(d) show the components $g_{xx}$, $g_{yy}$ and $g_{xy}$ of the quantum metric tensor, respectively, and (e) shows the Berry curvature $\mathfrak{B}_{xy}$. Colorscale units are square meters (m$^2$). Results of the quantum metric tensor are the same for the TM mode, and the Berry curvature of that mode differs by a minus sign from the TE result.}
	\end{figure*}

	To calculate the QGT, we first define a grid in a square portion of $k$-space that is centered at the $\Gamma$-point, and well within the first Brillouin zone, see upper panel of Fig.~\ref{fig:qgt_tmatrix}(a). The total size of the discretized surface is $10^{4}\times10^{4}$ m$^{-2}$, with a maximum grid spacing of $10$ m$^{-1}$ in both $k_{x}$ and $k_{y}$ directions. Then, we systematically calculate the eigenmodes at every point of the grid, using the band tracking procedure described in Section~\ref{sec:tmatrix_eigenf}, and focus on the TE and TM modes studied in Figs.~\ref{fig:tmatrix_bands} and~\ref{fig:losses_eigenvecs}. In Figs.~\ref{fig:losses_eigenvecs}(d),(e), we found that the dominant contributions to the eigenvectors are the in-plane dipolar coefficients $f_{21\pm 1}$, followed by the in-plane quadrupolar terms $f_{22\pm 2}$, independently of the path followed in $k-$space. We use this to identify our modes and ensure that the correct bands are followed. Moreover, we demand continuity of the dispersion energies with the $\mathbf{k}$-vector.
	
	According to Eq.~\eqref{def:berryphase}, the Berry phase can be calculated within closed loops connecting neighboring points of the grid in the upper panel of Fig.~\ref{fig:qgt_tmatrix}(a). From that calculation, taking the limit of a small loop, the Berry curvature may be obtained using Eq.~\eqref{berryphase_surface}. Here, however, we use a different method, namely we calculate numerical derivatives of the eigenfunctions explicitly at each point of the grid, see lower panel of Fig.~\ref{fig:qgt_tmatrix}(a). We set four additional data points (red crosses in Fig.~\ref{fig:qgt_tmatrix}(a)) to control the step lengths $\Delta k_{x,y}$ and achieve convergence of the numerical values of the derivatives $|\partial_{k_{x,y}} u_{\mathbf{k}}\rangle$ without increasing the number of grid points, thus speeding up the calculations. The derivatives are calculated as follows: 
	\begin{equation}\label{def:vswf_derivative}
		\dfrac{\partial|u_{\mathbf{k}}\rangle}{\partial k_{i}}\approx\dfrac{|u_{\mathbf{k}+\Delta k_{i}\hat{\mathbf{k}}_{i}}\rangle-|u_{\mathbf{k}-\Delta k_{i}\hat{\mathbf{k}}_{i}}\rangle}{2\Delta k_{i}}+\mathcal{O}(\Delta k_{i}^{2}),
	\end{equation}
	where $i=x,y$ and as indicated, the convergence of the derivative estimate is quadratic with the step size $\Delta k_{i}$. This estimate based on the Taylor expansion of $|u_{\mathbf{k}}\rangle$ is more efficient than linear convergence from usual approximations for right or left derivatives: $\partial_{k_{i}}|u_{\mathbf{k}}\rangle\approx(|u_{\mathbf{k}+\Delta k_{i}\hat{\mathbf{k}}_{i}}\rangle-|u_{\mathbf{k}}\rangle)/\Delta k_{i}+\mathcal{O}(\Delta k_{i})$. Overall, direct access to the numerical value of the derivatives $|\partial_{k_{x,y}} u_{\mathbf{k}}\rangle$ at each point of the grid $\mathbf{k}_{j}$ allows for a straightforward calculation of the QGT, providing all the components of both the quantum metric and the Berry curvature tensors simultaneously. For this purpose, we also define the inner product in the space of the VSWFs for two generic vectors $|u_{\mathbf{k}}\rangle$ and $|v_{\mathbf{k}}\rangle$ as follows:
	\begin{align}\label{def:vswf_innerproduct}
		\langle u_{\mathbf{k}}| v_{\mathbf{k}}\rangle&=\left(f^{*}_{21-1},f^{*}_{210},f^{*}_{211},\ldots\right)
		\begin{pmatrix}
			f'_{21-1} \\
			f'_{210} \\
			f'_{211} \\
			\vdots
		\end{pmatrix}\\
		&=f^{*}_{21-1}f'_{21-1}+f^{*}_{210}f'_{210}+f^{*}_{211}f'_{211}+\ldots\nonumber
	\end{align}
	
	With the considerations above, we insert definitions \eqref{def:vswf_derivative} and \eqref{def:vswf_innerproduct} into Eq.~\eqref{def:qgt} for the QGT, and calculate systematically the distribution of quantum metric and Berry curvature around the $\Gamma-$point. As we restrict our study to the two-dimensional $k-$space, we consider only the $g_{xx}$, $g_{yy}$, $g_{xy}$ and $g_{yx}$ components of the quantum metric tensor, and the components $\mathfrak{B}_{xy}$ and $\mathfrak{B}_{yx}$ of the Berry curvature. Other components are trivially zero in every point of the reciprocal space.
	
	Our results reveal a highly non-trivial profile in the quantum metric components and also in the Berry curvature, with an interesting non-zero distribution close to the diagonals of the first Brillouin zone, see the colorscale maps in Figs.~\ref{fig:qgt_tmatrix}(b)-(e). In particular, we find that the components $g_{xx}$ and $g_{yy}$ of the quantum metric (Figs.~\ref{fig:qgt_tmatrix}(b),(c)) are equal with positive contributions along both high-symmetry axes $k_{x}=\pm k_{y}$ in the analyzed area of $k-$space. On the other hand, the components $g_{xy}=g_{yx}$ (Fig.~\ref{fig:qgt_tmatrix}(d)) display very similar values to the other components of the quantum metric, but with a negative sign along the main diagonal of the Brillouin zone.
	
	These findings reflect on the symmetry of the square lattice since the diagonals of the Brillouin zone $k_{x}=\pm k_{y}$ are high-symmetry axes. However, we note that all the components of the quantum metric are zero in the rest of the analyzed area of $k$-space, including the other two high symmetry axes, $k_{x}=0$ and $k_{y}=0$, of this system. As is noticed in Ref.~\cite{CuerdaPRL2023}, the empty lattice bands along the diagonal $k_{x}=k_{y}$ show degeneracies of the TE and TM modes while in Figs.~\ref{fig:tmatrix_bands} and \ref{fig:losses_eigenvecs} these degeneracies are removed to yield a band splitting of the SLR modes, both in the real and in the imaginary part of the eigenfrequency.~This, in turn, results in abrupt changes in the phase of the corresponding eigenvectors close to the diagonal (e.g. see Figs.~\ref{fig:losses_eigenvecs}(d),(e), close to the point $B$ in $k-$space), and dramatically affects the eigenfunctions and the corresponding derivatives involved in the calculation of the QGT, see Eqs.~\eqref{def:quantummetric}, \eqref{def:berrycurvature} and \eqref{def:qgt}. The non-trivial features at the diagonals of the first Brillouin zone in Figs.~\ref{fig:qgt_tmatrix}(b)-(d) are thus a consequence of the band splitting and loss behavior of the SLR modes described in Figs.~\ref{fig:tmatrix_bands} and \ref{fig:losses_eigenvecs}.
	
	Of special interest is the emergence of non-zero Berry curvature close to the diagonals, since our work is concerned with a simple, symmetric square lattice where one would not expect it. The presence of Berry curvature in the first Brillouin zone is linked to the Chern number~\cite{Vanderbilt2018}:
	\begin{equation}\label{def:chern_number}
		C_{n}=\dfrac{1}{2\pi}\int_{BZ}d^{2}\mathbf{k}\mathbf{\mathfrak{B}}_{xy}^{n}(\mathbf{k}),
	\end{equation}
	where $n$ is the band index according to Eq.~\eqref{def:berrycurvature}. In contrast to the quantum metric components in Figs.~\ref{fig:qgt_tmatrix}(b)-(d) that display a symmetric distribution with respect to the diagonals of the Brillouin zone, the Berry curvature in Fig.~\ref{fig:qgt_tmatrix}(e) is antisymmetric, with contributions of equal magnitude and opposite sign at each side of all the diagonals. Hence the net contribution from each diagonal is zero, and the integral (\ref{def:chern_number}) over the analyzed area of $k-$space is canceled out. This is expected to extend to the rest of the Brillouin zone, and yield a zero Chern number for the square lattice case that we consider.
	
	Further analysis of the square lattice symmetry allows insight into the patterns in Figs.~\ref{fig:qgt_tmatrix}(b)-(e). We invoke here the relevant components of the quantum metric:
	\begin{align}
		g_{xx}&=\Re\{\langle\partial_{k_{x}}u_{\mathbf{k}}|\partial_{k_{x}}u_{\mathbf{k}}\rangle-\langle\partial_{k_{x}}u_{\mathbf{k}}|u_{\mathbf{k}}\rangle\langle u_{\mathbf{k}}|\partial_{k_{x}}u_{\mathbf{k}}\rangle\}\label{qm_xx},\\
		g_{yy}&=\Re\{\langle\partial_{k_{y}}u_{\mathbf{k}}|\partial_{k_{y}}u_{\mathbf{k}}\rangle-\langle\partial_{k_{y}}u_{\mathbf{k}}|u_{\mathbf{k}}\rangle\langle u_{\mathbf{k}}|\partial_{k_{y}}u_{\mathbf{k}}\rangle\}\label{qm_yy},\\
		g_{xy}&=\Re\{\langle\partial_{k_{x}}u_{\mathbf{k}}|\partial_{k_{y}}u_{\mathbf{k}}\rangle-\langle\partial_{k_{x}}u_{\mathbf{k}}|u_{\mathbf{k}}\rangle\langle u_{\mathbf{k}}|\partial_{k_{y}}u_{\mathbf{k}}\rangle\}\label{qm_xy},\\
		g_{yx}&=\Re\{\langle\partial_{k_{y}}u_{\mathbf{k}}|\partial_{k_{x}}u_{\mathbf{k}}\rangle-\langle\partial_{k_{y}}u_{\mathbf{k}}|u_{\mathbf{k}}\rangle\langle u_{\mathbf{k}}|\partial_{k_{x}}u_{\mathbf{k}}\rangle\}\label{qm_yx},
	\end{align}
	and explain the symmetric contributions of the quantum metric with respect to the diagonals of the Brillouin zone in Figs.~\ref{fig:qgt_tmatrix}(b)-(d). Upon a  mirror operation $\sigma_{v}$ with respect to the main diagonal $k_{x}=k_{y}$, the Brillouin zone remains invariant with the change of coordinates $k_{x}\rightarrow k_{y}$ and $k_{y}\rightarrow k_{x}$. Introducing this transformation in the derivatives of Eqs.~\eqref{qm_xx}-\eqref{qm_yx}, we obtain that $g_{ij}^{n}\rightarrow g_{ji}^{n}$ which results in an equal contribution on both sides of the diagonal since the quantum metric tensor is symmetric by definition: $g_{ji}^{n}=g_{ij}^{n}$. On the other hand, the $C_{4}$ symmetry with respect to the out-of-plane axis $\mathbf{k}_{z}$ allows four rotations, each with an angle of $m\pi/2$ where $m=1,\ldots,4$, that leave the Brillouin zone invariant. Therefore, the result of the quantum metric components in the first quadrant $k_{x,y}>0$ is extended to the rest of the Brillouin zone. For instance, a $\pi/2$ rotation from the first quadrant involves the change of coordinates $k_{x}\rightarrow k_{y}$ and $k_{y}\rightarrow-k_{x}$, introducing a minus sign in the derivative $\partial_{k_{x}}$ and thus in the components $g_{xy}$ and $g_{yx}$ in Eqs.~\eqref{qm_xy},\eqref{qm_yx} while leaving $g_{xx}$ and $g_{yy}$ in Eqs.~\eqref{qm_xx},\eqref{qm_yy} unchanged. This explains that while the quantum metric components $g_{xx}$ and $g_{yy}$ have the same magnitude and sign close to the diagonals in Figs.~\ref{fig:qgt_tmatrix}(b),(c), the components $g_{xy} = g_{yx}$ in Fig.~\ref{fig:qgt_tmatrix}(d) display equal magnitudes with an extra (-1) factor per each $\pi/2$ rotation.
	
	A similar analysis may be carried out for the Berry curvature in Fig.~\ref{fig:qgt_tmatrix}(e), whose expression reads as follows:
	\begin{equation}\label{bcxy}
		\mathfrak{B}_{xy}=i\left(\langle\partial_{k_{x}}u_{\mathbf{k}}|\partial_{k_{y}}u_{\mathbf{k}}\rangle-\langle\partial_{k_{y}}u_{\mathbf{k}}|\partial_{k_{x}}u_{\mathbf{k}}\rangle\right).
	\end{equation}
	By the same reasoning as above, it is found that the mirror operation $\sigma_{v}$ reverses the sign of $\mathfrak{B}_{xy}$ in Eq.~\eqref{bcxy}, leading to an antisymmetric distribution of Berry curvature along the main diagonal of the Brillouin zone. In addition, successive $\pi/2$ rotations leave Eq.~\eqref{bcxy} unchanged, hence the antisymmetric pattern found in the first quadrant ($k_{x,y}>0$) is extrapolated by rotations to the rest of the Brillouin zone. We note that the behavior in Fig.~\ref{fig:qgt_tmatrix}(e) is in accordance with the pseudovector nature of the Berry curvature \cite{Vanderbilt2018}.~Namely, one may define a pseudovector $\mathfrak{B}_{k}$ related to the Berry curvature tensor in Eq.~\eqref{def:berrycurvature} as follows: $\mathfrak{B}_{k}\equiv\epsilon_{ijk}\mathfrak{B}_{ij}$. The components of this pseudovector are given by the curl of the Berry connection: $\boldsymbol{\mathfrak{B}}=\nabla_{\mathbf{k}}\times\mathbf{A}$, where $\boldsymbol{\mathfrak{B}}=(\mathfrak{B}_{yz},\mathfrak{B}_{zx},\mathfrak{B}_{xy})$. As we restrict to the two-dimensional $k-$space, the pseudovector only has one non-zero component in the out-of-plane direction: $\boldsymbol{\mathfrak{B}}=\mathfrak{B}_{xy}\hat{\mathbf{k}}_{z}$.~The properties of a generic pseudovector $\mathbf{v}$ are such that it will transform as $\mathbf{v}'=(\det R)(R\mathbf{v})$ for a rotation $R$, which is either proper (e.g. rotations around an axis) with $\det R=1$, or improper (such as the mirror operation $\sigma_v$ considered above) with $\det R=-1$. Thus a mirror operation reverses the direction of the Berry curvature $\boldsymbol{\mathfrak{B}}'=-\mathfrak{B}_{xy}\hat{\mathbf{k}}_{z}$ with equal magnitude, explaining the change of sign at the diagonals of the Brillouin zone in Fig.~\ref{fig:qgt_tmatrix}(e), and rotations around the $\mathbf{k}_{z}$-axis leave the Berry curvature invariant.
	
	Overall, the quantum metric and Berry curvature found close to the $\Gamma$-point in Figs.~\ref{fig:qgt_tmatrix}(b)-(e) for a square plasmonic lattice are apparently unique to systems with long-range radiative interactions. Previous studies have identified non-zero QGT components around high-symmetry or other specific points in $k-$space \cite{HaldanePRL1988,HaldanePRL2008,RaghuPRA2008,GianfrateNature2020}, generally in tight-binding or continuum systems. Here, in contrast, we find that the non-zero contributions to the QGT are displayed along entire high-symmetry lines of the Brillouin zone, owing to the band structure of the system that is dominated by diffraction, see Ref.~\cite{CuerdaPRL2023}. As discussed above and in Ref.~\cite{CuerdaPRL2023}, the quantum metric is caused by the splitting in the real part of the eigenfrequency (band energy) and the Berry curvature strongly depends on the imaginary part of the splitting (losses), which explains the difference in magnitude between both.
	
	Importantly, we have found a Berry curvature that is non-zero at the local level, even for the highly symmetric, apparently trivial geometry of a square lattice. Such behavior, as discussed in Section~\ref{sec:tmatrix_eigenf}, points to a non-Hermitian origin since dissipation in the nanoparticles may break the time-reversal symmetry and could be linked with the band splitting along the diagonals of the Brillouin zone.
	
	\section{Loss dependence of the QGT}\label{sec:lossdepQGT}
	
\begin{figure}[b!]
		\includegraphics[width=0.97\columnwidth]{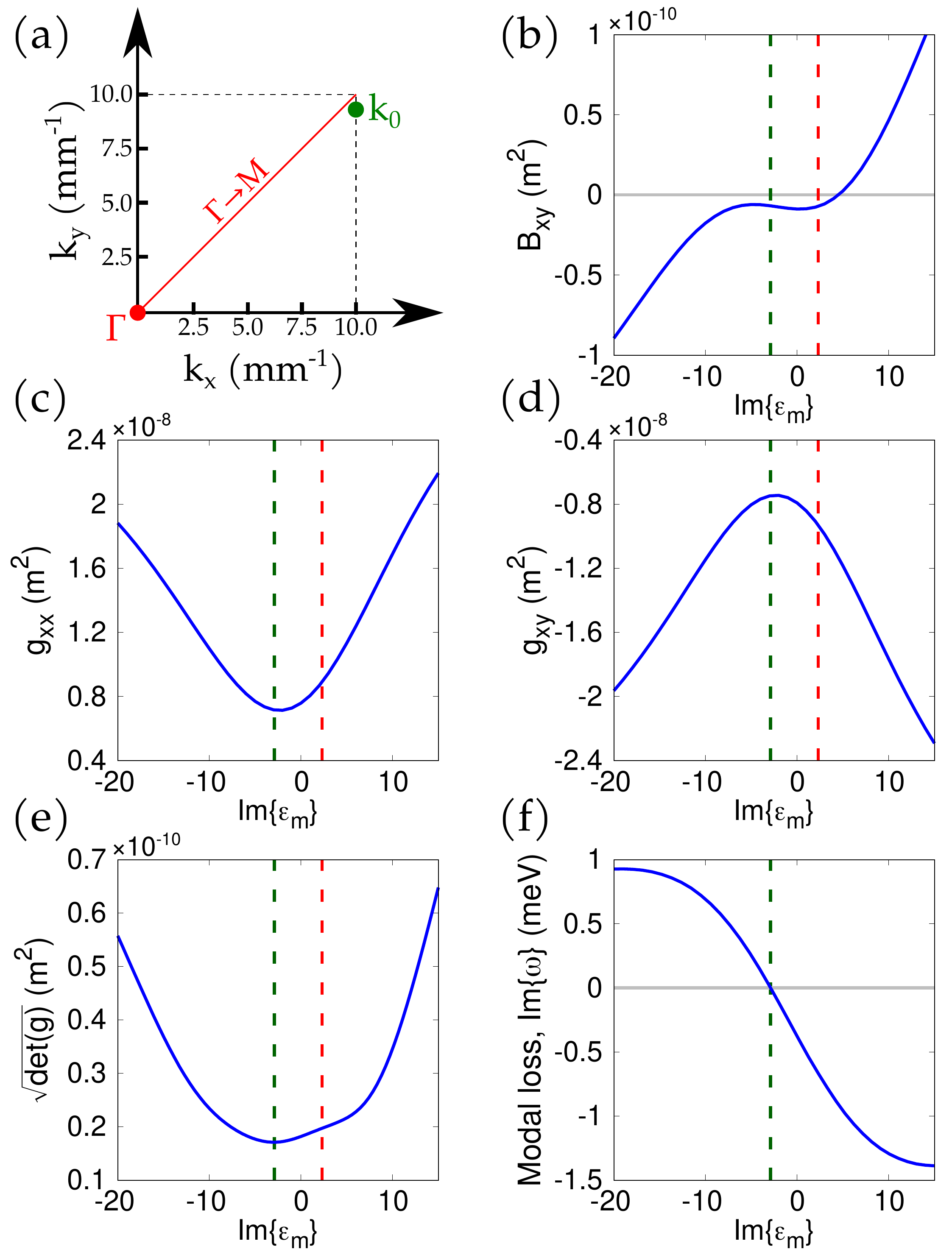}
		\caption{\label{fig:lossdep_qgt} Dependence of the QGT components (quantum metric tensor and Berry curvature) on dissipation losses. (a) We compute the quantum metric and Berry curvature at the point $\mathbf{k}_{0}=(10,9.5)$ mm$^{-1}$ in $k-$space (green point), close to the diagonal $\Gamma-M$ of the first Brillouin zone where non-zero components of the QGT are found in Fig.~\ref{fig:qgt_tmatrix}. (b) Berry curvature $\mathfrak{B}_{xy}$ at the point $\mathbf{k}_{0}$, calculated from several simulations analogous to those in Fig.~\ref{fig:qgt_tmatrix}.~The imaginary part of the permittivity $\epsilon_{m}$ characterizing the metallic nanoparticles is manually varied in each simulation, while the real part is kept constant at the physical value $\Re \epsilon_{m}\approx-26.3$. The physical value of $\Im\epsilon_{m}\approx 2.3$ corresponding to Fig.~\ref{fig:qgt_tmatrix} is marked as a red dashed line. In our convention, $\Im\epsilon_{m}>0$ corresponds to a lossy material, and $\Im\epsilon_{m}<0$ introduces gain in the nanoparticles. (c)-(e) Components $g_{xx}$ and $g_{xy}$ of the quantum metric, and square root of the determinant $\textrm{det}(g)=g_{xx}g_{yy}-g_{xy}g_{yx}$, respectively, calculated at the point $\mathbf{k}_{0}$ for varying $\Im\epsilon_{m}$ as described above.~For every value of $\Im\epsilon_{m}$, we found that $g_{yy}\approx g_{xx}$ and $g_{yx}\approx g_{xy}$ as in Fig.~\ref{fig:qgt_tmatrix}, hence these components are not shown. We have found that $\sqrt{\det g}>0.5|\mathfrak{B}_{xy}|$, as it should by definition \cite{OzawaPRB2021}, for every of $\Im\epsilon_{m}$ shown here. (f) Modal loss of the analyzed TE mode as a function of $\Im\epsilon_{m}$. Due to non-negligible radiative losses, the modal loss becomes zero at $\Im\epsilon_{m}\approx-2.9$ (green dashed line in panels (b)-(f)). This explains the asymmetry with respect to the value $\Im\epsilon_{m}=0$ (where only ohmic losses are compensated) in those panels. Results for the TM mode are the same as presented here, with an extra minus sign in the Berry curvature $\mathfrak{B}_{xy}$.} 
	\end{figure}

	We now address the intriguing question of why the Berry curvature becomes locally non-zero in a trivial lattice geometry with no apparent symmetry breaking. For that purpose, we clarify the role of losses in our results, and carry out additional simulations of the quantum metric and Berry curvature while varying the imaginary part of the metal permittivity $\epsilon_{m}$ corresponding to the ohmic losses, see Fig.~\ref{fig:lossdep_qgt}. By keeping the real part fixed to the physical value $\Re\epsilon_{m}\approx-26.3$ of Au, we do not aim to model any other realistic materials. For illustration, we study the evolution of the QGT components at a single point $\mathbf{k}_{0}$ of the Brillouin zone close to the diagonals where the quantum geometric features in Figs.~\ref{fig:qgt_tmatrix}(b)-(e) displayed maximum values (see Fig.~\ref{fig:lossdep_qgt}(a)).
	
	Fig.~\ref{fig:lossdep_qgt}(b) shows that the Berry curvature $\mathfrak{B}_{xy}$ at the point $\mathbf{k}_{0}$ depends highly on the dissipation losses. In particular, manually increasing the value of the metal permittivity beyond $\Im\epsilon_{m}\gtrsim7$ results in a linear increase of the Berry curvature, and a similar dependence is obtained below $\Im\epsilon_{m}\lesssim-9$ (for which gain instead of loss is numerically introduced to the involved TE mode) with opposite sign. The Berry curvature is negative but very small for values around $\Im\epsilon_{m}\approx 0$, including the physical value $\Im\epsilon_{m}\approx2.3$ of realistic lossy nanoparticles. Overall, the presence of local Berry curvature in $k-$space correlates well with the existence of losses (or gain) in the system. The non-trivial emergence of Berry curvature in our square plasmonic lattice is thus a non-Hermitian effect, produced by the time-reversal symmetry breaking through dissipation of the system, rather than induced by the lattice symmetry or its breaking.~Further evidence of this is provided in Ref.~\cite{CuerdaPRL2023}, where setting zero losses in the two-band model therein leads to a zero Berry curvature in every point of the Brillouin zone.
	
	The quantum metric components also exhibit an interesting dependence on ohmic losses, as their absolute values become larger for increased losses or gain, with a nearly symmetric distribution around values of $\Im\epsilon_{m}$ corresponding to low-loss particles, see Figs.~\ref{fig:lossdep_qgt}(c),(d). The slight asymmetry with respect to gain and loss points out that also other effects than dissipation affect the quantum geometric properties of the realistic system: we address these and offer additional physical insight with the simple two-band model in Ref.~\cite{CuerdaPRL2023}. We note that the quantum metric is positive semidefinite for every studied value of the ohmic losses, as it should,  see Fig.~\ref{fig:lossdep_qgt}(e).~We also find that additional radiative losses inherent to sufficiently large plasmonic nanoparticles introduce an asymmetry in the loss dependence of Figs.~\ref{fig:lossdep_qgt}(b)-(e) with respect to the value $\Im\epsilon_{m}=0$ (for which ohmic losses are neglected). Fig.~\ref{fig:lossdep_qgt}(f) reveals that both the radiative and ohmic losses of the analyzed TE mode are fully compensated for $\Im\epsilon_{m}\approx-2.9$ (green dashed line in Figs.~\ref{fig:lossdep_qgt}(b)-(f)), that coincides with the change of trend in Fig.~\ref{fig:lossdep_qgt}(b), and with the peak values in Figs.~\ref{fig:lossdep_qgt}(c),(d). We thus confirm the combined influence of both radiative and dissipative losses in the quantum geometric tensor components of a plasmonic lattice.

	\section{\label{sec:conclusions}Discussion and conclusions}
	We have studied the quantum geometric tensor, comprising the quantum metric and the Berry curvature, in a square plasmonic lattice. The band structure of the system was investigated using a T-matrix method that combines the optical properties of individual metallic nanoparticles with the long-range radiative interactions between them. This method includes the dissipative losses inherent to metals, enabling access to both the band energies and modal losses. By numerically tracking a closed path in reciprocal space, with trajectories parallel to the high-symmetry axes of the system, a band splitting was found along the diagonal of the first Brillouin zone. In addition, we calculated the quantum geometric tensor explicitly, with the complex-valued eigenvectors provided by the T-matrix approach, and found non-trivial distributions of quantum metric and Berry curvature in all the diagonals of the Brillouin zone.
	
	The results here presented clarify, with microscopical simulations, the origin of pseudospin orbit-coupling in a square plasmonic lattice. In particular, we have shown numerically that a TE-TM band splitting emerges at the diagonal of the Brillouin zone, but it is absent in the empty lattice results presented in  Ref.~\cite{CuerdaPRL2023}. Therefore, the band splitting at the diagonals is caused by the presence of the nanoparticles, and not by the lattice symmetry. This is in agreement with the statements and with the experimental observations in Ref.~\cite{CuerdaPRL2023}, and the polarization properties shown therein verify the TE and TM nature of the involved modes at the diagonal. Our numerical simulations presented here provide realistic values of the band splitting that are utilized as an input for the two band model in Ref.~\cite{CuerdaPRL2023}. The results of all the QGT components show excellent qualitative agreement with those in Ref.~\cite{CuerdaPRL2023}. Our results, in combination with the two-band model with $\Omega_{x,y}\neq 0$ and $\Omega_{z}=0$, hint the possible existence of Fermi arcs and exceptional points due to the non-hermiticity~\cite{BergholtzRevModPhys2021}.
	
	Our numerical simulations also yield a non-zero antisymmetric Berry curvature in the diagonals of the Brillouin zone. We have found that the Berry curvature correlates with loss (or gain) in the system, corroborating its non-Hermitian origin. The Berry curvature is non-zero at the local level, but its distribution strongly suggests that the Chern number is zero. Thus, in this case, the breaking of time-reversal symmetry generates non-trivial quantum geometry locally but without topologically non-trivial behavior. Overall, our results provide a microscopical understanding to the pioneering observation of non-Hermitian Berry curvature in a plasmonic lattice~\cite{CuerdaPRL2023}. They provide a basis for extending the studies of topological photonics in plasmonic lattices and similar long-range coupled platforms, to include effects of various geometric, particle shape, magnetic field, and gain. With such symmetry breaking mechanisms, a rich variety of novel topological and quantum geometric phenomena can be expected.   
	
	\begin{acknowledgments}
		
		We acknowledge useful discussions with Mikko Rosenberg and Marek Ne{\v{c}}ada. This work was supported
		by the Academy of Finland under Project No.~349313, Project No.~318937
		(PROFI), and the Academy of Finland Flagship Programme in
		Photonics Research and Innovation (PREIN) Project
		No. 320167. J.C.~acknowledges former support by the Academy of Finland under project No.~325608. J.M.T.~acknowledges financial support by the Magnus Ehrnrooth Foundation. Part of the research was performed at the OtaNano Nanofab cleanroom (Micronova Nanofabrication Centre), supported by Aalto University.
	\end{acknowledgments}
	
	\appendix
	
	\section{T-matrix formalism and simulations}\label{appendix:tmatrix}
	
	We use a frequency-domain T-matrix method to formulate the multiple-scattering problem of a lattice of metallic nanoparticles interacting with light. We model a two-dimensional lattice in the three-dimensional space, embedded in a dielectric with refractive index $n_{h}$. The optical response of the metallic (Au) nanoparticles is introduced with a Drude-Lorentz model. The T-matrix method numerically solves the wave equation in this system, by expanding the incident and the scattered electric fields in a basis of vector spherical wavefunctions (VSWFs)~\cite{Necada2021,Mishchenko2000}:
	\begin{align}
		\mathbf{E}(\omega,\mathbf{r})&=\sum_{\tau=1,2}\sum_{l=1}^{\infty}\sum_{m=-l}^{l}(a_{\tau lm}\mathbf{v}_{\tau lm}(\kappa\mathbf{r})\nonumber\\
		&\qquad\qquad\qquad\qquad\qquad+f_{\tau lm}\mathbf{u}_{\tau lm}(\kappa\mathbf{r}))\label{Efield_expansion_into_VSWF}.
	\end{align}
	Here, $\mathbf{v}_{\tau lm}(\kappa \mathbf{r})$ are the so-called \emph{regular} VSWFs, whereas $\mathbf{u}_{\tau lm}(\kappa \mathbf{r})$ are named \emph{outgoing} VSWFs. The subindex $\tau$ refers to whether the VSWF is \emph{electric} or \emph{magnetic}-type; $l$ is the multipole degree and $m$ is the multipole azimuthal number. Ref. \cite{Necada2021} contains complete expressions of the VSWFs.
	
	Without the presence of the scatterer, the regular VSWFs would suffice to form a basis that solves the Maxwell equations for a homogeneous dielectric background; hence we identify the $a_{\tau lm}$ coefficients as those corresponding to the incident electric field: $\mathbf{E}_{inc}(\omega,\mathbf{r})=\sum_{\tau lm}a_{\tau lm}\mathbf{v}_{\tau lm}(\kappa\mathbf{r})$. Similarly, the $f_{\tau lm}$ coefficients correspond to the scattered electric field $\mathbf{E}_{scat}(\omega,\mathbf{r})$. Both sets of coefficients are connected by the \emph{transition} \mbox{(T-)matrix }\cite{Kristensson2016}:
	\begin{equation}\label{def:tmatrix}
		f_{\tau lm}=\sum_{\tau'l'm'}T_{\tau lm}^{\tau'l'm'}(\omega)a_{\tau'l'm'},
	\end{equation}
	where we have highlighted that the T-matrix depends on frequency.
	
	Moving to the problem of scatterers arranged in a square lattice, we may similarly expand the electric field as in Eq.~\eqref{Efield_expansion_into_VSWF} for each scatterer $p$ placed in position $\mathbf{r}_{p}$ on the two-dimensional plane of the lattice:
	\begin{align}
		\mathbf{E}_{p}(\omega,\mathbf{r})=\sum_{\tau=1,2}\sum_{l=1}^{\infty}\sum_{m=-l}^{l}(a_{p,\tau lm}\mathbf{v}_{\tau lm}(\kappa(\mathbf{r}-\mathbf{r}_{p}))\nonumber\\
		+f_{p,\tau lm}\mathbf{u}_{\tau lm}(\kappa(\mathbf{r}-\mathbf{r}_{p}))).
	\end{align}
	Due to the presence of other scatterers, the $a_{p,\tau lm}$ coefficients contain not only the contribution of the incident electric field, but also from scattered fields by other lattice sites. We may re-expand scattered fields at other lattice sites $\mathbf{r}_{q}\neq \mathbf{r}_{p}$ into the basis of regular VSWFs $\mathbf{v}_{\tau lm}$ around the site $p$, by means of a translation operator $S_{p\leftarrow q}$ \cite{Necada2021}. As a result, the expression for the coefficients $a_{p,\tau lm}$ now reads as follows:
	\begin{equation}\label{coefs_a_taulm}
		a_{p,\tau lm}=\Tilde{a}_{p,\tau lm}+\sum_{q\neq p}S_{p\leftarrow q}f_{q,\tau lm}.
	\end{equation}
	Multiplying Eq.~\eqref{coefs_a_taulm} by $f_{q,\tau lm}$ from the left, and using Eq.~\eqref{def:tmatrix}, we find:
	\begin{equation}\label{tmatrixproblem_generic_scatterer_distr}
		f_{p}-T_{p}\sum_{q\neq p}S_{p\leftarrow q}f_{q}=T_{p}\Tilde{a}_{p},
	\end{equation}
	where we have dropped the subindices $\tau$, $l$, and $m$ for clarity.
	
	Eq.~\eqref{tmatrixproblem_generic_scatterer_distr} formally solves the scattering problem for a set of particles in positions $\mathbf{r}_{p}$ on a two-dimensional plane. For a Bravais lattice of scatterers, the positions of the particles are labelled by $p\rightarrow\mathbf{n},\alpha$; where $\mathbf{r}_{\mathbf{n},\alpha}=\mathbf{R}_{\mathbf{n}}+\mathbf{r}_{\alpha}$, and $\mathbf{R}_{\mathbf{n}}=n_{1}\mathbf{a}_{1}+n_{2}\mathbf{a}_{2}$, where $\mathbf{a}_{1,2}$ are the vectors of the Bravais lattice and $n_{1,2}$ are integers. We similarly relabel $q\rightarrow\mathbf{m},\beta$ and redefine the origin of coordinates using periodicity; therefore $S_{\mathbf{n},\alpha\leftarrow\mathbf{m},\beta}=S_{\mathbf{0},\alpha\leftarrow\mathbf{m-n},\beta}$. Further, we use the Bloch theorem to write $\Tilde{a}_{\mathbf{n},\alpha}=\Tilde{a}_{\mathbf{0},\alpha}(\mathbf{k})e^{i\mathbf{k}\cdot\mathbf{R}_{n}}$, and similarly for $f_{\mathbf{n},\alpha}$. Hence Eq.~\eqref{tmatrixproblem_generic_scatterer_distr} is rewritten as
	\begin{equation}\label{tmatrix_periodic_scatterer_distr_longform}
		f_{\mathbf{0},\alpha}(\mathbf{k})-T_{\alpha}\sum_{\beta}W_{\alpha\beta}(\mathbf{k})f_{\mathbf{0,\beta}}(\mathbf{k})=T_{\alpha}\Tilde{a}_{\mathbf{0},\alpha}(\mathbf{k}),
	\end{equation}
	where the term $W_{\alpha\beta}(\mathbf{k})\equiv\sum_{\mathbf{m}}(1-\delta_{\alpha\beta}\delta_{\mathbf{m0}})S_{\mathbf{0},\alpha\leftarrow\mathbf{m},\beta}e^{i\mathbf{k}\cdot\mathbf{R_{m}}}$ computes the infinite lattice (Ewald) sum, and hence implements the long-range radiative interactions that characterize our lattice \cite{Necada2021}. Writing Eq.~\eqref{tmatrix_periodic_scatterer_distr_longform} in the matrix form, we end up with
	\begin{equation}\label{tmatrix_periodic_scatterer_distr_shortform}
		(I-TW)f_{\mathbf{0}}(\mathbf{k})=T\Tilde{a}_{\mathbf{0}}(\mathbf{k}).
	\end{equation}
	Expression (\ref{tmatrix_periodic_scatterer_distr_shortform}) solves for the scattering problem and has a clear interpretation: the right-hand side contains a forcing vector with coefficients $\Tilde{a}_{\mathbf{0},\tau lm}$ of the incident electric field; the left-hand side accounts for both the single particle response via the $T$ matrix and the contribution from the lattice geometry via $W$. The coefficients $f_{\mathbf{0},\tau lm}(\mathbf{k})$ are the unknowns that solve the scattering problem.
	If the forcing term is set to zero $T\Tilde{a}_{\mathbf{0},\tau lm}=\mathbf{0}$, Eq.~\eqref{tmatrix_periodic_scatterer_distr_shortform} turns into an eigenmode and eigenvalue problem. The QPMS suite utilized in this paper \cite{QPMScode} provides an implementation of the Beyn's algorithm \cite{Beyn2012} that solves this eigenvalue problem and provides the eigenfunctions for the direct calculation of the QGT using Eqs.~\eqref{def:vswf_derivative},~\eqref{def:vswf_innerproduct}, and \eqref{def:qgt}.

	
	%
	
\end{document}